\documentclass[10pt,twocolumn,letterpaper]{article}

\usepackage{cvpr}
\usepackage{times}
\usepackage{epsfig}
\usepackage{graphicx}
\usepackage{amsmath}
\usepackage{amssymb}
\usepackage{algorithm} 
\usepackage{algorithmic} 
\usepackage{caption}
\usepackage{subcaption}
\usepackage{xcolor} 

\usepackage[breaklinks=true,bookmarks=false]{hyperref}

\cvprfinalcopy 

\ifcvprfinal\fi
\begin{document}

\title{3-D Context Entropy Model for Improved Practical Image Compression}

\author{Zongyu Guo, Yaojun Wu, Runsen Feng, Zhizheng Zhang,  Zhibo Chen\thanks{Zongyu Guo and Yaojun Wu contribute equally to this work. Zhibo Chen is the corresponding author.} \\
\textit{CAS Key Laboratory of Technology in Geo-spatial Information Processing and Application System}\\
\textit{University of Science and Technology of China}\\
{ \small \tt \{guozy,yaojunwu,fengruns,zhizheng\}@mail.ustc.edu.cn, chenzhibo@ustc.edu.cn}}

\maketitle

\begin{abstract}
In this paper, we present our image compression framework designed for CLIC 2020 competition. Our method is based on Variational AutoEncoder (VAE) architecture which is strengthened with residual structures. In short, we make three noteworthy improvements here. First, we propose a 3-D context entropy model which can take advantage of known latent representation in current spatial locations for better entropy estimation. Second, a light-weighted residual structure is adopted for feature learning during entropy estimation. Finally, an effective training strategy is introduced for practical adaptation with different resolutions. Experiment results indicate our image compression method achieves 0.9775 MS-SSIM on CLIC validation set and 0.9809 MS-SSIM on test set.
\end{abstract}

\section{Introduction}

Image compression is a ubiquitous technique in the digital age. Traditional image compression standards take years to develop a new generation. With the rapid development of Deep Neural Networks (DNNs), learning-based image compression method presently is attractive and achieves some promising breakthroughs \cite{chen2019learning,guo2019deep,he2019beyond}. Early learning-based method \cite{toderici2017full} is based on RNN and supports coding scalability. However, image compression is a rate-distortion trade-off game and such RNN-related work cannot directly optimize the rate during network training.

Recently, most learning-based image compression approaches are based on VAE architecture, where rate $R$ and distortion $D$ are jointly optimized in an end-to-end manner \cite{balle2016end}. Ball{\'e} \etal \cite{balle2018variational} propose a hyperprior entropy model, which parameterizes the latent distribution and predicts their standard deviations as Gaussian Scale Model (GSM). After that, \cite{lee2018context} and \cite{minnen2018joint} introduce context entropy model to utilize adjacent known regions for better parameter estimation and improve original GSM to Single Gaussian Model (SGM). Recent works \cite{lee2019hybrid,cheng2020learned} further suggest a more generalized format to predict the distribution of latent representation, \ie, Gaussian Mixture Model (GMM). GMM theoretically is able to approximate arbitrary continuous probability distribution. Those impressive improvements mentioned above mainly concentrate on the hyperprior model for parameter estimation. Additionally, the backbone network can also be enhanced with some techniques such as attention mechanism \cite{zhou2019end,cheng2020learned} and post-processing network \cite{lee2019hybrid}.

In this paper, motivated by the aforementioned methods, we build our image compression network for CLIC 2020 low rate track and highlight three main improvements. First, we propose a 3-D context entropy model which divides latent representations into two groups across channels. This 3-D context model can better extract correlations of latent features which are in the same spatial location but vary in channel. Second, a residual structure is adopted to refine the estimated entropy parameters. The designed residual parameter estimation (RPE) module efficiently cooperates with the 3-D context model thanks to the light-weighted but effective structure. Third, a novel training strategy is employed for practical image compression. We know that due to the downsampling layer in network, learning-based codec usually requires the input to have an integer-multiple resolution of values such as 32 or 64. Consequently, when dealing with such images with different resolutions, we should first conduct padding process. This may lead to unnecessary bit waste on padded areas. The proposed training strategy enables network to adapt to different padding situations in the time of training.

In CLIC 2020 low rate track, our team IMCL\_IMG \_MSSSIM got 0.9775 and 0.9809 MS-SSIM during the validation phase and test phase respectively.

\section{Method}

\begin{figure*}[t]
  \centering
  \includegraphics[scale=0.36, clip, trim=1.5cm 38cm 9cm 1.5cm]{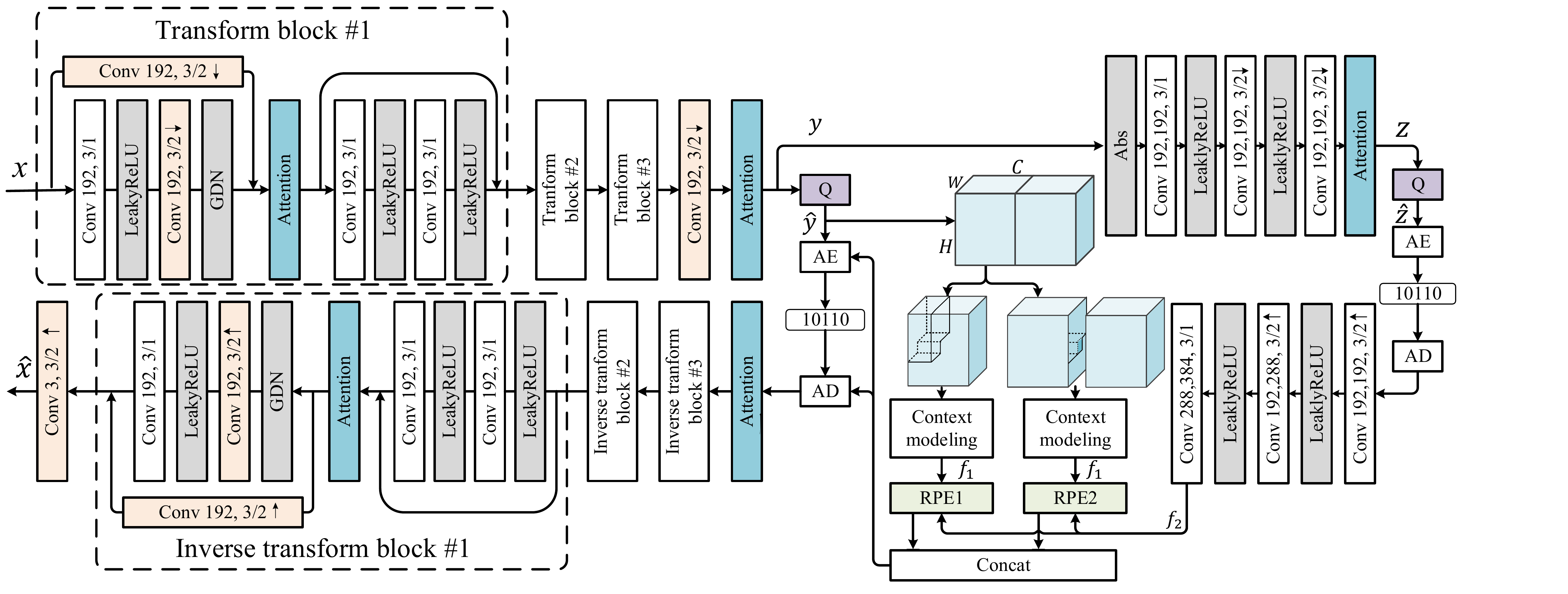}
  \caption{The overall framework of our image compression model. The context entropy model is a 3-D version.}
  \label{figure1}
\end{figure*}

\subsection{The overall framework}

Before introducing our proposed new techniques, we first present our overall framework, which is shown in Figure~\ref{figure1}. Similar to previous work \cite{minnen2018joint}, this pipeline can be divided into three parts: an analysis transform encoder, a synthesis transform decoder, and a hyperprior entropy model (including a hyper encoder and a hyper decoder). The hyperprior part will be discussed later. This backbone network is an improved version based on \cite{cheng2020learned}. Specifically, the raw input image $x$ will be transformed to the latent features $y$, which will be quantized to $\hat{y}$ and then decoded to reconstructed image $\hat{x}$. The analysis transform encoder contains three transform blocks, each of which is made of a residual downsampling layer, an attention layer and a residual enhancement layer. After three transform blocks, there are a downsampling convolution layer and an attention layer to increase receptive field. The architecture of synthesis decoder is symmetric, \ie, an attention layer, three inverse transform blocks and an additional upsampling layer.

Compare with the baseline network \cite{cheng2020learned}, we modify their model with several extra attention modules in the encoder side, which has no increasing complexity for decoding (two in analysis transform encoder and one in hyper encoder). Besides, GRDN \cite{kim2019grdn}, a post-processing network recommended in \cite{lee2019hybrid}, is adopted following the main compression network to further enhance image quality, which is omitted in Figure \ref{figure1}.


\subsection{3-D context entropy model}

As a part of hyperprior model, context entropy model was first proposed in \cite{minnen2018joint} and \cite{lee2018context}. This context model is autoregressive over latents and is usually implemented in the format of $5 \times 5$ mask convolution \cite{minnen2018joint}. Such context entropy model plays an important role for the estimation of feature parameters though it would increase decoding time complexity dramatically. 

The mask convolution layer in previous context model can effectively capture spatial correlations to predict current pixel, which is similar to classical intra prediction. Our experiments indicate that not only spatial redundancy can be eliminated, there also exists channel-wise redundancy, even though Generalized Divisive Normalization (GDN) is proved to well Gaussianize features in the channel direction. 

Assuming we are predicting current latent representation $y$, its location is $[i,j,k]$, where $i$ and $j$ are the coordinate of height and width and $k$ is the channel location index. While original 2-D context model concentrates on the left and up features $\hat{y}_{i-h,j-w}$, the proposed 3-D context model further leverages known (decoded) features in current spatial location, \ie, $\hat{y}_{i,j,k-c}$. Ideally, different channel requires different mask convolution in our 3-D context model, \eg, feature in the first channel can be predicted only with the up and left features but feature in the last channel can be predicted with those known features in current spatial location. However, this ideal situation will complicate model because in this case, every channel should have its own parameter estimation module. Therefore, we finally choose to compromise which divides all channels into two groups. Each group has its own weights of mask convolution and now there are two independent parameter estimation modules for those two groups. As shown in Figure \ref{figure1}, the first group is predicted as usual but the second group can be predicted based on the first group. 

The proposed 3-D context model enables the sequential decoding process to be more sequential. We have tried to divide channels into more groups, which was found to improve little. This 3-D context model is analogous to the conditional RGB prediction model in PixelCNN \cite{oord2016pixel}. The difference is that here are more channels rather than only three in PixelCNN and we divide these channels into two groups for simplification.

\subsection{Residual parameter estimation module}

\begin{figure}[t]
 \centering
 \includegraphics[scale=0.285, clip, trim=18.2cm 14cm 10cm 36cm]{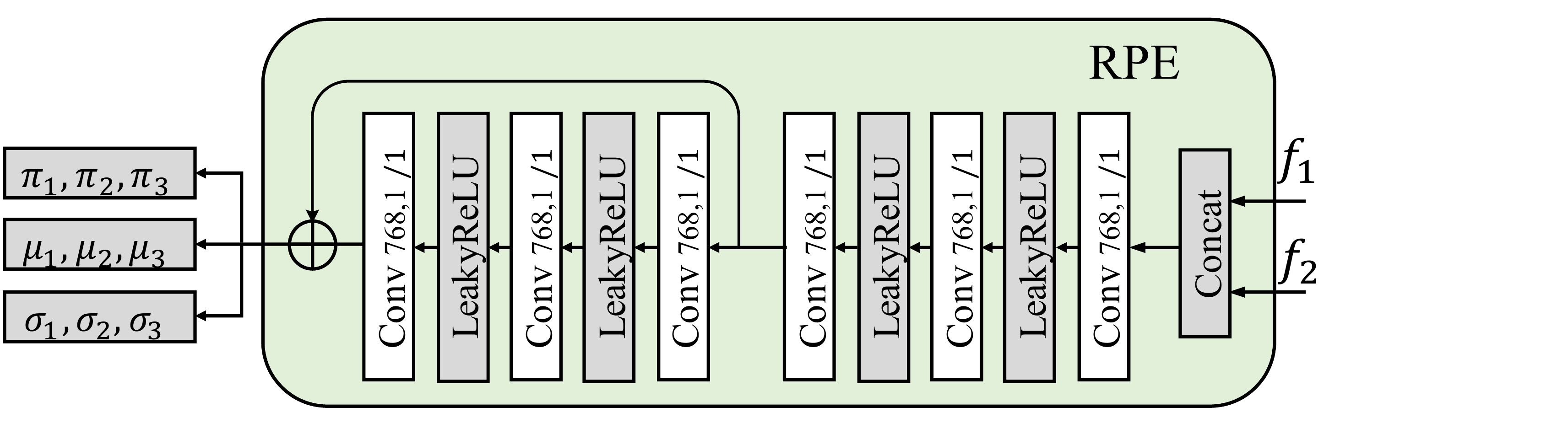}
 \caption{Residual parameter estimation module. There is only one residual connection and thus the RPE module is relatively lighted-weighted but effective.}
 \label{figure2}
\end{figure}

Cheng \etal \cite{cheng2019deep} comprehensively discuss the residual architecture for image compression. As they shown, residual structures in analysis transform and synthesis transform obviously strengthen the capability of network. Motivated by this, we think the entropy parameter estimation module can also be enhanced with the help of residual structure. 

As shown in Figure \ref{figure1}, after obtaining context feature $f_1$ and hyper feature $f_2$ reconstructed from $\hat{y}$ and $\hat{z}$, we employ a residual parameter estimation (RPE) module to estimate the probability distribution of $\hat{y}$. As mentioned before, the distribution of latent features is modeled as Gaussian Mixture Model (GMM) following \cite{lee2019hybrid,cheng2020learned}, \ie, 
\begin{equation}
p(\hat{y}) \sim \sum_{k=1}^{K} \pi_{k}  \mathcal N(\mu_{k}, \sigma^2_{k}). 
\end{equation}
In our experiments, we find that $K=3$ is enough to accurately estimate the distribution of latent representations. The structure of RPE module is presented in Figure \ref{figure2}. There are three $1 \times 1$ convolution layers to process the concatenation of $f_1$ and $f_2$. Then there follows a residual component which also contains three $1 \times 1$ layers. Such $1 \times 1$ residual convolution layers, which can also be regarded as fully connected layers, mainly work for features across the entire channels instead of spatial features. Therefore, it will not influence the sequential decoding process.


Lee \etal \cite{lee2019hybrid} propose a Model Parameter Refinement Module (MPRM) to cooperate with global context. Our designed residual entropy parameter module is partially different from theirs because here we only have one residual block, which is effective and light-weighted. The moderate parameter number is also advantageous for the 3-D context model because the 3-D context model here doubles the whole parameter number of entropy estimation module.

\subsection{More practical image compression}

Practical image compression codec requires to handle those images with different resolutions. It is problematic even for traditional block-based image compression methods, \eg, VTM (VVC test model) \cite{VTMref} would first change image resolution to an integer multiple of 8. Considering learning-based image compression methods, this problem is always more serious because there are many downsampling layers in network which would cause resolution inconsistency after inversion. A conventional solution would be extra padding process before encoding. However, learning-based network is usually trained with full cropped patches such as $256 \times 256$ patches. As a result, the network cannot handle these padded image properly in practical applications and then performance usually drops.

In our framework, there are totally six downsampling layers and thus input images should have an integer multiple resolution of $2^6 = 64$. First we note that experiments prove that zero-padding is optimal than other padding methods such as reflection-padding. Here we propose a strategy to enable network to adapt to the padding effects during training. The pseudo code of proposed algorithm is as following in Algorithm \ref{alg1}.

\renewcommand{\algorithmicrequire}{\textbf{Input:}}
\renewcommand{\algorithmicensure}{\textbf{Output:}}

\begin{algorithm} 
\caption{Training strategy for practical compression} 
\label{alg1} 
\begin{algorithmic}[1] 
\REQUIRE A mini-batch data $x$ randomly cropped from training dataset, the shape of which is $[B,C,H,W]$

\STATE \textit{Flag} $\leftarrow$ random sample $\in \{0,1\} $.
\IF{\textit{Flag} is 0}
\STATE Normally optimize your network.
\ELSE
\STATE Randomly get the padding size for current batch.
\STATE Select $h_{pad} \in [0, P_1]$, $w_{pad} \in [0, P_2]$.
\STATE Zero-pad input $x$ right and down. Then its shape is $[B,C,H + h_{pad},W + w_{pad}]$.
\STATE Crop $x$ to simulate the real input image after padding: $x=x[:,:,h_{pad}:, w_{pad}:]$.
\STATE Calculate pixel number $(H-h_{pad}) \times (W-w_{pad})$ to obtain actual bitrate $R$ of current batch. Then aggregate the distortion loss $D$ which only covers unpadding area.
\STATE Optimize your network.
\ENDIF
\end{algorithmic}
\end{algorithm}

In this algorithm, $P_1$ and $P_2$ are given upper bound to control the padding size during training. In our experiments, considering that input patch is $256 \times 256$ (H=W=256), we empirically set $P_1$ = $P_2$ = 20. In short, we want to enable network to have access to padded images even if those images are imitated by manually crafted padding. Randomly choosing padding size will help network adapt to different images with different padding situation. This training strategy is verified to largely improve the performance in CLIC validation dataset (the actual required bitrate decreases).

\section{Implementation details}

We train our network with $256 \times 256$ patches randomly cropped from CLIC training set, DIV2K and Flickr 2K dataset, which has the same setting as \cite{zhou2019end}. We divide the training period into three stages. We first train our main compression network without post-processing. Then we fix the parameters in the main compression network and train corresponding post-processing module GRDN \cite{kim2019grdn}. At the last step, we jointly optimized the whole pipeline to achieve the best results. Notably, the proposed training strategy for padding effect is applied only at the third stage. At different training stages, we all take a learning rate decay strategy, \ie, $lr=1e-4$ in the initial 300,000 iterations and $lr=1e-5$ for the rest 300,000 iterations. We train the network on two RTX 2080 Ti GPUs when batch size is set to 8.

Due to the limit of 0.15 bpp in CLIC competition, we train different models for different compression ratios. As usual, the loss function is $\mathcal L =R+\lambda D$, where $D=1-msssim$. Note that the loss function is modified when we employ the proposed training strategy for padding at the third training stage. Considering that we optimize for MS-SSIM, we select appropriate $\lambda$ value ranging from 10 to 24. Table \ref{table1} shows the results of our methods including single model and multiple models. However, it seems that four models have little improvement compared with two models, which may imply that our rate control strategy is not satisfactory and has room to improve. Our final submitted version is this four-model codec, which achieves 0.9775 MS-SSIM score for validation and 0.9809 MS-SSIM for test.

\begin{table}[t]
\centering
\resizebox{\columnwidth}{!}{
\begin{tabular}{|c|c|c|c|}
\hline
Model & MSSSIM & PSNR & BPP\\
\hline
Single model ($\lambda=16$) & 0.97812 & 30.30 & 0.1548 \\
\hline
Two models ($\lambda=\{12,16\}$) & 0.97753 & 30.19 & 0.1499 \\
\hline
Four models ($\lambda$ from $[10,24]$) & 0.97754 & 30.19 & 0.1499 \\
\hline
\end{tabular}}
\caption{Performance on CLIC 2020 validation dataset. Optimized for MS-SSIM.\label{table1}}
\end{table}

\section{Conclusion}

In this paper, we introduce our image compression framework used in CLIC 2020 competition. Three useful techniques are adopted. First we improve the conventional 2-D context entropy model to a 3-D format which can better utilize decoded features in current spatial location. Second, the parameter estimation module is enhanced with a light-weighted residual structure. Lastly, we propose a training strategy to handle those images with different resolutions in real application. This training strategy is simple but effective to alleviate the bit waste due to preliminary padding. As shown on the leaderboard, our team IMCL\_IMG\_MSSSIM got the second place in terms of MS-SSIM in the validation phase. 
In the future, we will pay more attention to more practical image compression methods, \ie, lighter, faster and more robust DNN-based codec.

\section*{Acknowledgment}
This work was supported in part by NSFC under Grant U1908209, 61632001 and the National Key Research and Development Program of China 2018AAA0101400.

{\footnotesize
\bibliographystyle{ieee_fullname}
\bibliography{egbib}
}

\end{document}